\documentclass[final,12pt]{elsarticle}

\usepackage{amssymb}
\usepackage{xcolor}
\usepackage{tikz}
\usetikzlibrary{shapes,decorations}
\usepackage{hyperref}
\usepackage{siunitx}
\usepackage{caption}
\usepackage{subcaption}
\usepackage{xurl}
\usepackage{mathtools}
\usepackage{comment}
\usepackage{geometry}

\usepackage{lineno}
\usepackage[ruled,vlined,linesnumbered]{algorithm2e}
\usepackage{algpseudocode} 
\usepackage{listings}
\definecolor{codegreen}{rgb}{0,0.6,0}
\definecolor{codegray}{rgb}{0.5,0.5,0.5}
\definecolor{codepurple}{rgb}{0.58,0,0.82}
\definecolor{backcolour}{rgb}{0.97,0.97,0.92}
\lstdefinestyle{mystyle}{
    backgroundcolor=\color{backcolour},   
    commentstyle=\color{codegreen},
    basicstyle = \ttfamily\scriptsize,
    keywordstyle=\color{magenta},
    numberstyle=\tiny\color{codegray},
    stringstyle=\color{codepurple},
    breakatwhitespace=false,         
    breaklines=true,                 
    captionpos=b,                    
    keepspaces=true,                    
    numbersep=3pt,                  
    showspaces=false,                
    showstringspaces=false,
    showtabs=false,                  
    tabsize=2
}
\journal{journal}
\begin{document}

\begin{frontmatter}

\title{Synthetic Porous Microstructures: Automatic Design, Simulation, and Permeability Analysis}

\author[1,2,3]{Thomas Lavigne}
\author[1]{Camilo Andrés Suarez Afanador}
\author[1]{Anas Obeidat}
\author[1]{Stéphane Urcun}
\cortext[*]{Corresponding author}
\ead{stephane.urcun@uni.lu}

\affiliation[1]{organization={Institute of Computational Engineering, Department of Engineering, University of Luxembourg},
addressline={6, avenue de la Fonte},
city={Esch-sur-Alzette},
postcode={L-4364},
country={Luxembourg}}

\affiliation[2]{organization={Arts et Metiers Institute of Technology, IBHGC},
addressline={151 bd de l’hopital},
city={Paris},
postcode={75013},
country={France}}

\affiliation[3]{organization={Arts et Metiers Institute of Technology, Univ. of Bordeaux, CNRS, Bordeaux INP, INRAE, I2M Bordeaux},
addressline={Avenue d'Aquitaine},
city={Pessac},
postcode={33607},
country={France}}

\begin{abstract}
This study introduces an open-source computational framework for the generation and permeability evaluation of synthetic porous media. The proposed methodology integrates crystallographic and meshing tools to construct controlled microstructures with tunable porosity, facilitating seamless transitions from geometric modelling to computational domains for numerical simulations. The generated structures are analysed through fluid-structure interaction (FSI) simulations, leveraging the Entropically Damped Artificial Compressibility (EDAC) formulation in conjunction with the Discretization-Corrected Particle Strength Exchange (DC-PSE) method.

A novel approach for the numerical estimation of the macroscopic permeability tensor is presented, employing a stochastic upscaling technique inspired by the volume averaging method. To validate the framework, we investigate the transition between creeping and non-Darcy flow regimes through parametric permeability studies, utilizing a one-dimensional approach for practical benchmarking.

The results establish a foundation for experimental validation and provide insights into the customised design of porous structures for engineering and biomedical applications, offering a versatile tool for research in fluid transport and porous media mechanics.
\end{abstract}


\begin{highlights}
\item Open-source framework for synthetic porous medium generation.
\item Particle method to evaluate the permeability of the medium.
\item Simulated permeability compared to the experiment.
\end{highlights}

\begin{keyword}
Porous media \sep Permeability \sep Computational Fluid Dynamics \sep Homogenisation \sep Fluid-Structure Interaction
\end{keyword}

\end{frontmatter}



\section{Introduction}
\label{sec:Introduction}

Porous media, whether natural or synthetic, encompass a diverse range of materials, including natural substances such as soil, rock, sponges, and wood, as well as biological tissues such as trabecular bone and menisci. In addition, engineered porous structures, such as foams, concrete, electrospun fibre networks, titanium-printed scaffolds for tissue engineering and paper-based materials, have expanded the functional applications of porous media~\cite{Gibson:1999, Ashby:2000, Banhart:2001}. These materials share a distinctive combination of advantageous physical properties, including low weight, high permeability, large surface area to volume ratio, initial stiffness, energy absorption capacity and capillary-driven liquid attraction~\cite{Boomsma:2002, Lage:1996, Bhattacharya:2002}.

Because of these properties, porous media are employed in diverse applications, such as shock absorption, load-bearing components, heat exchangers, catalysts, fluid mixing, and solid lubrication. Fundamentally, all porous materials consist of a solid framework interspersed with interconnected voids that enable fluid transport~\cite{cowin1985}. The interaction between the solid structure and the fluid that flows through it often leads to complex behaviours, particularly when considering how fluid dynamics influence the mechanical properties of the medium~\cite{boutin1991}. These phenomena, collectively referred to as fluid-structure interactions (FSIs), are of critical importance in multiple fields, including biomechanics, materials science, and engineering~\cite{biot1956}.

While natural porous materials exhibit remarkable multifunctionality, their intricate microstructures pose significant challenges for analysis and simulation. First, accurately modelling their complex, heterogeneous architecture requires computationally intensive simulations for both fluid and solid mechanics. Second, their microstructural variability, which differs spatially within the same sample and between different samples, introduces uncertainties that complicate deterministic modelling and prediction.

This inherent randomness can be a limiting factor in industrial applications. For instance, open-cell metal foams offer substantial advantages in automotive engineering, yet inconsistencies in their microstructure, even under identical manufacturing conditions, hinder their widespread adoption. To overcome these limitations, synthetic porous media have emerged as an alternative, offering the advantage of controlled fabrication processes~\cite{bear1972}. By designing synthetic porous materials with customised pore size, distribution, and connectivity, researchers and industry professionals can optimise these structures for specific applications. As a result, synthetic porous media are being increasingly used in sectors such as water purification, petrochemistry, and acoustics~\cite{adler1992, Tien1993, Bear:1972, Allard:2009}.

In biomechanics, the variability in natural porous structures, such as knee menisci, complicates personalised medical treatments, as their microstructure differs between individuals. The ability to fabricate synthetic porous materials with precise and reproducible properties enhances the reliability of the investigation and ensures consistent quality in medical implants, thus improving patient outcomes~\cite{wu2016}. Furthermore, conducting \textit{in vivo} biomechanical experiments presents challenges related to sample geometry, boundary conditions, and mechanical response measurements. This introduces multiple sources of uncertainty, raising concerns about the robustness of the derived conclusions. A promising alternative involves the use of synthetic porous materials as standardised experimental models, combined with \textit{in silico} simulations. Since the physical domain of a synthetic porous medium, its geometry and porosity are well defined, initial constitutive parameters can be accurately determined, either through computational modelling or experimental validation. Furthermore, synthetic porous media facilitate the isolation of specific variables in scientific studies, a task often complicated by the inherent variability of natural samples. They can also be engineered to replicate pathological conditions, supporting targeted therapeutic research and intervention strategies~\cite{martinez2007}.

With respect to the synthetic generation of porous media, previous studies have focused on deterministic replication of porous structures using idealised repetitive units. These models include uniformly arranged sphere beds~\cite{Krishnan:2006}, cubic unit cells composed of 65 square cylinders~\cite{Plessis:1994, Boomsma:2001}, and diamond-configured cells with vertical and horizontally connected micro-cylinders~\cite{Bai:2011}. A representative elementary volume (REV) provides a more realistic depiction of microstructural porosity~\cite{Boomsma:2002, Plessis:1994, Hill:1963}, requiring fewer adjustments compared to idealised models~\cite{Krishnan:2006, Plessis:1994, Boomsma:2001}.

Various computational methods have been developed to analyse fluid transport and heat transfer in porous media. \citet{Aarnes:2008} introduced mixed multiscale finite element methods (MsFEM) to solve stochastic flow equations with permeability-dependent basis functions, enabling large-scale solutions. \citet{Ganapathysubramanian:2009} proposed a multiscale stochastic model that captures unpredictable subgrid-scale permeability variations using a reduced-order approach based on statistical data. \citet{Biswal:2011} developed a multiscale framework to characterise porous media at varying levels of detail, accommodating a range of pore sizes. More recently, \citet{Vasic:2022} investigated the impact of the micropore structure on the performance of porous material using higher-order modelling techniques. In the context of subsurface flow, \citet{Dostert:2008} proposed an efficient sampling method to quantify uncertainty in permeability distributions, leveraging coarse-scale models and sparse interpolation to approximate posterior distributions.

Building upon these advancements, the present study aims to develop a streamlined methodology for the generation of random porous microstructures using a combination of open-source Python, POV-Ray SDL, and C++ libraries. This approach takes advantage of highly optimised computational functions, allowing for efficient automation of the generation process. Furthermore, we establish a framework for estimating macroscopic permeability tensors by conducting fluid-structure interaction simulations, following the methodology proposed by \citet{whitaker1998method} and refined by \citet{Scandelli2022}.



\section{Methodology}
\label{sec:mat&meth}
\subsection{Generation of synthetic samples}
\label{sec:geom}

\begin{figure}[ht!]
\centering
\includegraphics[width=\textwidth]{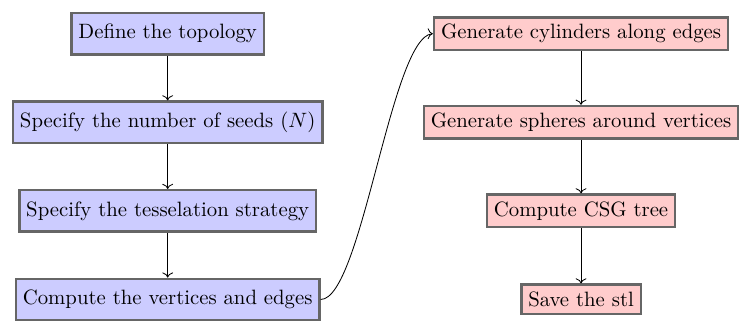}
\caption{Graphical Representation of the Workflow for Porous Scaffold Generation. Blue elements indicate the Neper environment, while red elements represent the Pymesh environment.}
\label{fig:1}
\end{figure}

The pipeline developed for the generation of synthetic geometries, as illustrated in Figure \ref{fig:1}, leverages the open-source crystallographic software Neper~\cite{Quey2011} in combination with the open-source Python meshing library Pymesh~\cite{zhou2019pymesh}. This integrated approach ensures efficient and precise control over geometric and meshing parameters. The complete set of scripts and codes used for mesh generation is provided as supplementary material to facilitate reproducibility and further research.

The topology of the porous medium, \textit{i.e.}, the spatial domain occupied by the porous scaffold, was first defined (Algorithm \ref{algo1}). Standard geometric shapes such as cylinders, prisms, and tori are available as predefined options. However, more complex geometries can also be generated through transformations, including Boolean operations. In this study, the porous medium was confined within a cylindrical domain with a diameter of 24 \si{\milli\meter} and a height of 40 \si{\milli\meter}.

The macroscopic porosity ($\phi$) of the domain was determined by a single parameter: the number of seeds $N$, which corresponds to the target number of tessellation cells. To illustrate the functionality of the generation process, four cases with varying seed numbers were calculated: $N=1500,~2000,~2500,~3000$. In addition, a morphological rule governing the tessellation process was specified, allowing customisation according to statistical constraints, Voronoi tessellation principles, or other user-defined criteria. Depending on the chosen approach, the resulting scaffold could exhibit either a controlled, regular pore size or a more randomly distributed microstructure. In the present work, a logarithmic normal sphericity law was applied to ensure a relatively uniform pore-size distribution. Based on these parameters, a set of vertices and edges was generated and stored in a ``.tess" file, representing the tessellation output.

\begin{algorithm}[ht!]
\caption{\href{https://neper.info/}{NEPER}: Generate a domain and its corresponding tessellation}
\label{algo1}
\SetKwInOut{Input}{input}\SetKwInOut{Output}{output}

    \Input{Geometric specifications}
\Output{Tessellation file}
\BlankLine
    $N \gets \{1500,~2000,~2500,~3000\}$\;
    $L_z \gets 4.0\times10^{-2}\si{\meter}$\;
    $r \gets 1.2\times10^{-2}\si{\meter}$\;
    DOMAIN $\gets$ ``cylinder({$L_z$},~{$2*r$})"\;
    Law $\gets$ ``diameq:dirac(1),1-sphericity:lognormal(0.145,0.03)"\;
    \eIf {$\exists$ output.tess}{Pass}{output.tess $\gets$ neper -T -n \$N -morpho \$Law -domain \$DOMAIN}

\end{algorithm} 

After generation, the vertices and edges from the ``.tess" file were imported into the Pymesh environment (Algorithm \ref{algo2}). Then a cylindrical structure was assigned along each edge, with a fixed radius of 0.5 \si{\milli\meter}. The methodology can be extended to incorporate variable radii by defining a function that adjusts the radius based on the distance between the connected vertices, ensuring proportional scaling across the structure. To maintain a smooth and continuous scaffold geometry, spherical elements were added at the boundary vertices, preventing abrupt discontinuities. In addition, an external cylindrical tube was introduced, creating a 10 \si{\milli\meter} chamber on either side of the sample to define a confined region for fluid flow. Given the high number of generated elements, a Constructive Solid Geometry (CSG) tree was employed to perform Boolean operations, resulting in a well-defined and unified surface mesh.

\begin{algorithm}[ht!]
\caption{\href{https://pymesh.readthedocs.io/en/latest/}{Pymesh}: Create the porous scaffold}
\label{algo2}
\SetKwInOut{Input}{input}\SetKwInOut{Output}{output}

\Input{output.tess to load the vertices and edges (parsing)}
\Output{Mesh of the porous scaffold}
N $\gets$ \{1500, 2000, 2500, 3000\}\;
rad $\gets$ $5\times10^{-4}\si{\meter}$\;
list $\gets$ [~]\;
$i \gets$ 0\;
v, e $\gets$ M([x,y,z]), M([v(i), v(j)]); \Comment{vertices and edges}\;
\While {$i\leq$ len(e)}{
    cylinder $\gets$ pymesh.generate\_cylinder(v[e[0]], v[e[1]], radius, radius)\;
    list $\gets$ [list, cylinder]\;
    $i\gets i+1$;}
$i \gets$ 0\;
\While {$i\leq$ len(v)}{
\If{ v on boundary}{
    sphere $\gets$ pymesh.generate\_icosphere(radius, v[i])\;
    list $\gets$ [list, sphere];  }

$i\gets i+1$;  }
list $\gets$ [list, tube];        \Comment{tube generated with Pymesh}\;
mesh $\gets$ CSG tree; \Comment{boolean operations between all parts}
\end{algorithm}

Once the porous scaffold was created, it was saved as a ``.stl" file—whether generated via the aforementioned pipeline or obtained from the segmentation of a micro-CT image. This geometry was then processed using the numpy-stl library, which ensured the correct orientation of surface normals. A set of custom functions, based on the prior stl-to-voxel library~\cite{Pederkoff2015}, was implemented to convert the mesh into a voxel-based representation. Users can define the desired voxel resolution ($N_x \times N_y \times N_z$ voxels) or specify a fixed voxel size, depending on computational requirements.

\begin{algorithm}[ht!]
\caption{STL-2-Voxel: Transform the mesh.stl to a particle space}
\label{algo3}
\SetKwInOut{Input}{input}\SetKwInOut{Output}{output}

\Input{Mesh of the porous scaffold}
\Output{Particle space}
$x,~y,~z \gets 0,~0,~0$\;
min, max $\gets$ mesh spatial limits\;
$vx_{size}$ $\gets 1\times 10^{-3}$\;
$ROI$ $\gets$ max-min\;
$idx$ $\gets$ argmax(ROI)\;
$res$ $\gets int(\frac{ROI[idx]}{vx_{size}})$ \Comment{expected resolution}\;
$voxels,~scale,~shift \gets$ stltovoxel.convert(mesh, resolution=$res-1$)\;
\While {$z\leq voxels.shape[0]$}{
    \While {$y\leq voxels.shape[1]$}{
        \While {$x\leq voxels.shape[2]$}{
            \textbf{write}  $[x, y, z] / scale$  in coord.txt; \Comment{sphere center\;}
            \eIf {point $\notin$ confined well}{
               \textbf{write} 3 tag.txt; }{\textbf{write} voxels[z][y][x] tag.txt; }
        $x \gets x+1$;}
    $y \gets y+1$;}
$z \gets z+1$;}
\end{algorithm}

For synthetic porous media, additional considerations were necessary to ensure that the fluid remained within the confined region. Specifically, voxels located outside the external cylindrical boundary were removed to prevent undesired flow artifacts. This step was feasible due to the well-defined dimensions of the domain. A similar adjustment may be required when replicating an experimental set-up with an external boundary, but it can be omitted in other cases.

Following voxelisation, the voxel coordinates and their corresponding material tags were exported to separate files, facilitating domain creation in subsequent processing steps (Algorithm \ref{algo3}). The resulting microstructures are presented in the implementation section (Section \ref{sec:res}).

\subsection{Governing equations and mathematical modelling of fluid/structure interactions}
\label{sec: Math Modeling}
For clarity and convenience, we provide a brief overview of the fundamental concepts underlying the governing equations. Readers seeking a more comprehensive examination of the Entropy-Density Approximate Compressibility (EDAC) formulation in conjunction with the Discretization-Corrected Particle Strength Exchange (DC-PSE) method are encouraged to refer to~\cite{Singh:2023}.

\subsubsection{Entropically Damped Artificial Compressibility (EDAC) formulation}

In the field of simulating the incompressible Navier-Stokes equations, \cite{Clausen:2013} introduced the EDAC method, which allows for the explicit simulation of these equations. The EDAC formulation incorporates an additional governing equation for the evolution of pressure $p$ (Equation \ref{EDAC}), derived from the thermodynamic properties of the system while ensuring a fixed density $\rho$. This method demonstrates convergence to the incompressible Navier-Stokes (INS) equations at low Mach numbers and maintains consistency across a broad range of Reynolds numbers, from low to high. The momentum equation and the pressure evolution equation within the Eulerian frame of reference,

\begin{eqnarray}
\rho\frac{du_i}{dt} + u_j \frac{\partial u_i}{\partial x_j}&=&   - \frac{\partial p}{\partial x_i}  +  
                         \frac{\partial \tau_{ij}}{\partial x_j} 
\label{momentum-euler}
\\
\frac{d p}{dt}  +  u_i\frac{\partial p}{\partial x_i} &=&  -{c_s}^2 \rho_o   \frac{\partial u_i}{\partial x_i} + \nu \frac{\partial^2 p}{\partial x_i x_i}, 
\label{EDAC}
\end{eqnarray}
\begin{equation}
   \label{shear stress}
   \tau_{ij} = \mu 
      \left( \frac{\partial u_i}{\partial x_j} + 
             \frac{\partial u_j}{\partial x_i} - 
             \frac{2}{3} \delta_{ij} \frac{\partial u_k}{\partial x_k} 
      \right),
\end{equation}

where $u$ is the velocity vector field, $p$ is the pressure field, $t$ is time, $\tau$ is the shear stress, $\mu$ is the dynamic viscosity, and $c_{s}$ is the speed of sound.

Equation~\ref{momentum-euler} corresponds to the equation that governs momentum conservation, while Equation~\ref{EDAC} represents the EDAC formulation for the evolution of pressure. This formulation introduces entropy as a means of damping pressure oscillations.
A comprehensive derivation, along with the physical model of the EDAC formulation, is extensively detailed in Clausen's work~\cite{Clausen:2013}.
In summary, starting from the compressible Navier-Stokes equations, we derive the pressure evolution equation using mass conservation, entropy balance, and thermodynamics, introducing temperature as a variable and adding an entropy constraint. Clausen resolves the issue of density fluctuations by linking density directly to pressure and temperature, simplifying the model to explicitly relate temperature to pressure in Equation~\ref{EDAC}. Hence, allowing for explicit pressure evolution.

In this study, the EDAC equations are coupled with Brinkman penalisation, as explained in the reference~\cite{Obeidat:2019,Obeidat:2020}, for the simulation of fluid flow in intricate geometries. This approach involves the implicit penalisation of the computational domain, achieved by employing an indicator function $\chi$ that identifies the areas occupied by the solid geometry denoted $O$.
 \begin{equation}
 \label{chi-eq}
\chi(x) =\left\{\begin{matrix}
1 & \text{if} \, \, \, x \in O,\\ 
 0&  \text{otherwise}.& 
\end{matrix}\right.
 \end{equation}
A penalty term is added to the momentum equation (implicit penalisation). The penalized conservation of momentum equation is:
\begin{equation}
\label{p-momentum}
 \rho\frac{du_i}{dt} + u_j \frac{\partial u_i}{\partial x_j}= -\frac{\partial p}{\partial x_i}  +  
                         \frac{\partial \tau_{ij}}{\partial x_j} 
                         - \frac{\chi}{\eta}({u}_i - u_{(oq)i})
\end{equation}
Where, ${u}_{(oq)i}$ is the velocity of the solid body, $\eta = \alpha \phi$ is the normalized viscous permeability and $\phi$ is the porosity. Note that $0 < \phi < 1$ and $0 < \eta \ll 1$.
\subsubsection{The Discretization-Corrected Particle Strength Exchange (DC-PSE) formulation}

DC-PSE is a numerical technique utilised to discretise differential operators when dealing with unevenly distributed collocation points~\cite{Schrader:2010}. It initially emerged as an enhancement to the traditional Particle Strength Exchange (PSE) method~\cite{Degond:1989a, Eldredge:2002}, with the aim of decreasing the integration error when the coordinate points are placed irregularly. However, from a mathematical point of view, it essentially represents an extension of finite differences~\cite{Schrader:2010}. The group of collocation methods within the PSE/DC-PSE category employs mollification, utilising a symmetric smoothing kernel $\eta \epsilon()$, to approximate continuous functions $f_\epsilon(\vec{x})$ that are sufficiently smooth, as:
\begin{align}
		f(\vec{x}) \approx f_{\epsilon}(\vec{x})=\int_{\Omega} f(\vec{y}) \eta_{\epsilon}(\vec{x}-\vec{y}) \mathrm{d} \vec{y},
	\end{align}

where $\epsilon$ is the smoothing length or the width of the kernel.
In DC-PSE the linear system is solved locally to each particle in order to determine the kernel weights in a way that they locally satisfy the
discrete moment conditions to the desired order of convergence. This allows DC-PSE to directly satisfy the discrete-moment conditions on particle distribution rustling, avoiding quadrature error.

The most commonly used DC-PSE kernels are of the form

\begin{align}
\eta(\vec{x})=\left\{\begin{array}{ll}\sum_{i, j}^{i+j<r+m+n} a_{i, j} x^{i} y^{j} e^{-x^{2}-y^{2}} & \sqrt{x^{2}+y^{2}}<r_{c} \\ 0 & \text { otherwise, }\end{array}\right.
\end{align}

where the polynomial coefficients $a_{i,j}$ are determined from the discrete moment conditions 

\begin{align}
Z^{i, j}\left(\vec{x}_{p}\right)=\left\{\begin{array}{ll}i ! j !(-1)^{i+j} & i=m, j=n \\ 0 & \alpha_{\min }<i+j<r+m+n \\ <\infty & \text { otherwise. }\end{array}\right.
\end{align}

$\alpha_{min}$ is $0$ for odd and $1$ for even operators, and the discrete moments $Z^{i, j}$ are defined as

\begin{align}
Z^{i, j}\left(\vec{x}_{p}\right)=\sum_{\vec{x}_{q} \in \mathcal{N}\left(\vec{x}_{p}\right)} \frac{\left(x_{p}-x_{q}\right)^{i}\left(y_{p}-y_{q}\right)^{j}}{\epsilon\left(\vec{x}_{p}\right)^{i+j}} \eta\left(\frac{\vec{x}_{p}-\vec{x}_{q}}{\epsilon\left(\vec{x}_{p}\right)}\right).
\end{align}

This not only leads to operator discretizations that are consistent on (almost\footnote{DC-PSE fails on particle distributions where particle positions in the neighbourhood are linearly dependent. In such cases, the linear system for the kernel weights does not have full rank and cannot be solved.}) all particle distributions, but also relaxes the overlap condition of PSE to the less restrictive requirement

\begin{align}
\frac{h\left(\vec{x}_{p}\right)}{\epsilon\left(\vec{x}_{p}\right)} \in \mathcal{O}(1),
\end{align}

that is, the ratio of the kernel width $\epsilon$ and the inter-particle spacing $h$ has to be bounded by an arbitrary constant as $h\to 0$.

\subsection{Numerical evaluation of the permeability of the medium}
\label{sec:eval_num}

\subsubsection{Theoretical hypotheses and validity criteria}

In this section, we analyse the existence of an effective permeability tensor $\mathbf{\Tilde{K}}$ for a non-periodic synthetic porous medium generated following the methodology presented in Section \ref{sec:geom}. The EDAC solver for DC-PSE~\cite{Singh:2023} is used to model the three-dimensional incompressible viscous fluid flow within the synthetic porous medium by solving the EDAC formulation as described in Section \ref{sec: Math Modeling}. This simulation scenario is in agreement with the formulation of the effective (or macroscopic) permeability tensor from \textit{The Method of Volume Averaging} \cite{whitaker1998method} which is the main reference work for further derivations. In the case of non-periodic microstructures, to determine the applicability of the above work in simulations, the recent work in \cite{Scandelli2022}  is used as a complementary reference, since it presents an extensive review of the literature, several application cases, and introduces a supplementary criterion verifying ideal conditions for such estimations. 

Within this framework, we assume that the observed medium can be macroscopically described by a Darcy relationship, which represents a linear dependency of the macroscopic physical fields, the spatial average of the velocity and the gradient of pressure distributions. This relation is expressed via the macroscopic Darcy's law obtained by the volume averaging method for non-slip condition on the interphase fluid/solid, as it is described in \cite{Howes1985, Whitaker1986, whitaker1998method}:
\begin{equation}
    \langle u \rangle =-\mathbf{\Tilde{K}}\frac{\nabla \langle p \rangle}{\mu \Phi} = -\mathbf{\Tilde{K}}\frac{\langle \nabla p \rangle}{\mu \Phi}
\end{equation}
Where $\langle f \rangle$ is the spatial average of a function $f(x)$ in a domain $\Omega$ defined as:
\begin{equation}
    \langle f \rangle=\frac{1}{||\Omega||}\int_{\Omega}f(x)d\Omega,
\end{equation}
and $u,p$ are the velocity and pressure fields respectively, $\mathbf{\Tilde{K}}$ the macroscopic permeability tensor, $\mu$ the dynamic viscosity, $\Phi$ the macroscopic porosity. 
For such relationships to hold, it is necessary to verify that the flow can be considered to be in the so-called ``creeping regime'', in which non-linear inertial effects are negligible (from a macroscopic point of view), and thus a linearity of the macroscopic fields can be considered. For this, the main parameter to consider is the Reynolds number ($Re$) and additionally the normalised vorticity criterion ($w^{*}$) introduced in \cite{Scandelli2022}. The explicit form of these two parameters is presented below:

\begin{equation}
    Re = \frac{\rho \langle \boldsymbol{u}\rangle l_c}{\mu}, \text{ and, } w^* = \frac{w l_c}{||\langle \boldsymbol{u} \rangle||} = \frac{(\nabla \times \boldsymbol{u})l_c}{||\langle \boldsymbol{u} \rangle||}.
\end{equation}

Following the methodology of \textit{volume averaging} \cite{whitaker1998method}, we consider a Representative Elementary Volume (REV) as a subset of the macro-domain holding the same porosity of the entire domain, which verifies the linearity requirements of Scandelli-Whitaker.
However, periodic boundary conditions are not valid in the studied cases as the random microstructures of Section \ref{sec:geom} are the geometrical support for the simulations. Therefore, we compute for each REV in a set of sample REVs the effective permeability tensor as follows: 

\begin{equation}
\langle u_i \rangle = -\frac{\kappa_{ij}(x_i)}{\mu \Phi} \nabla \langle p(x_i) \rangle
\end{equation}

Given that for each point of a REV we know the velocity and pressure field, we need to compute $\kappa_{ij}$ the effective permeability tensor of an intermediate scale created by the spatial distribution of REVs accounting for the different distributions of the microscopic fields, but to do this, we first have to take into account that the number of equations from the above linear system is less than the required for the calculation of the elements in this tensor. To be able to estimate a solution for such a system, we assume that there is no fracture in the domain and conservation of momentum that leads to symmetry and positive definite properties of the tensor \cite{pouya2002definition}.
The initial Darcy problem within the fluid phase (i.e. $\chi (x_i) = 0$) reads: 
\begin{equation}
\begin{split}
& \text{Given $\{u_i(x_i),p(x_i)\}$:}\\
& \text{Find $\kappa_{ij}(x_i)  \forall x_i \in \Omega_{REV}$,   } \langle u_i \rangle = -\frac{\kappa_{ij}(x_i)}{\mu \Phi}\nabla \langle p(x_i) \rangle, \chi (x_i) = 0\\
\end{split}
\end{equation}

To simplify the notation, let us introduce the average material pressure gradient vector $\langle p^*_{i}(x_i)\rangle = (\mu \Phi)^{-1}\langle \nabla p(x_i) \rangle$ to rewrite the above local Darcy law, 
\begin{equation}
    \langle u_i \rangle = \kappa_{ij} \langle p^*_{j} \rangle, 
\end{equation}
and, to ease the formulation we adopt a matrix-vector notation with vectors represented by the column matrix of dimensions $dim(u)=(3,1)$, and the local permeability tensor a square matrix of dimensions $dim(\boldsymbol{\kappa})=(3,3)$: 
\begin{equation}
    \langle \boldsymbol{u} \rangle = \boldsymbol{\kappa}\langle \boldsymbol{p}^* \rangle
\label{eq: Darcy problem}
\end{equation}

The effective permeability tensor $\boldsymbol{\kappa}$ being symmetric by definition (i.e. $\boldsymbol{\kappa} = \boldsymbol{\kappa}^{T}$) the number of independent coefficients to compute is 6, therefore the linear system of three equations relating the 6 coefficients is undetermined and the solution cannot be directly obtained. To overcome this issue, one should refer to \cite{Scandelli2022}, in which 3 compatible simulations are used to recover the elements of the effective permeability tensor (see also \cite{POUYA2002975, Lang2014}).

\paragraph{Vorticity criterion}
As mentioned above, to start the exploratory analysis of the existence of a macroscopic permeability tensor $\tilde{K}$, first, one must observe the distribution of the norm of the vorticity vector ($\| \boldsymbol{w} \|= \|\nabla \times \boldsymbol{u} \|$) that verifies the pertinence of such computation via the supplementary indicator of the presence of the creeping regime introduced in \cite{Scandelli2022}, that in the case of the reference article, is given by comparisons for different Reynolds numbers using a dimensionless vorticity (weighted by up-scaling parameters) with respect to a reference state, validating the Darcy-like flow, and therefore, verifying the validity of a macroscopic permeability tensor following \cite{whitaker1998method}. 

\paragraph{Parametric up-scaling} 
The following steps represent a deductive approach to identify the parameter $l_c$, a characteristic length that preserves the so-called representativeness of the macroscopic medium, describing a domain decomposition in which the elements are rich enough (from a statistical point of view) to be a simplified representation of the entire domain. Since the effective permeability tensor is a function of the porosity volume fraction, the macroscopic volume fraction $\phi$ is used here as a parameter to guide the choice of $l_c$ for the example microstructure. At this point, a first arbitrary choice is made; this is the shape of the subdomains. For the present study, cuboid subdomains are chosen to facilitate the numerical implementation since the subdomains are represented by structured meshes (as a consequence of the voxelisation process for simulations). First, the distribution of local porosity is observed for a set of test characteristic lengths, and the parameter is chosen such that the obtained distribution is a Gaussian-like distribution centred close to the expected volume fraction, and the dispersion around the mean is symmetric and of reasonable amplitude. The corresponding values are discussed in Section \ref{sec:res}.

\subsubsection{Stochastic approximation of the macroscopic effective permeability}
Following \cite{Scandelli2022}, we are able to compute an approximation of the effective permeability tensor $\boldsymbol{\kappa}$ of the ``intermediate" scale for a given REV, which in terms of the spatial distribution is characterised by its centroid. The domain decomposition led by $l_c$ ensures the representativeness of the medium, and thus a smooth decomposition is capable of considering every material point within the domain as a potential centroid of an REV. Moreover, such decomposition allows the REVs to overlap mapping in a smooth way the changes of the distributions of the ``moving" REV sampling. Thus, the macroscopic permeability tensor can be obtained via stochastic upscaling using Monte Carlo integration, which gives as a result an effective behaviour that takes into account the local porosity statistics of the random porous medium.

\section{Implementation}
\label{sec:res}

\subsection*{Sample Geometries generation}

\begin{figure}[ht!]
\centering
        \includegraphics[width=0.7\textwidth]{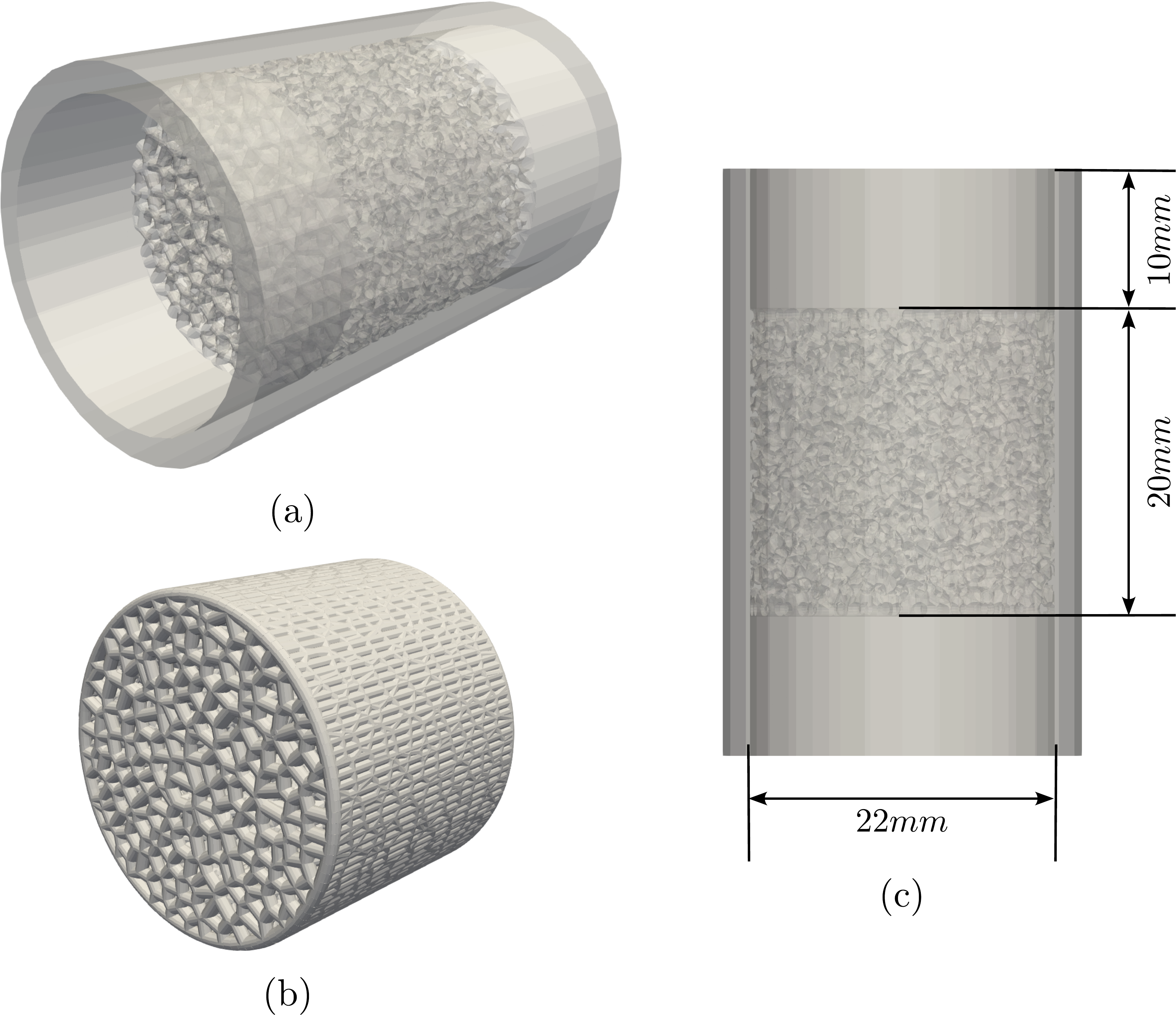}
        \caption{Sample Microstructure $N=3000$. (a) A 3D isometric representation of the generated STL geometry, (b) a closer view to the inner porous network inserted in the cylindrical domain for simulations, and (c) the dimensions of the confined well configuration for simulations on a 2D projection of the domain.}
    \label{fig: comp domain}
\end{figure}

\begin{figure}[ht!]
    \centering
    \includegraphics[width=0.9\textwidth]{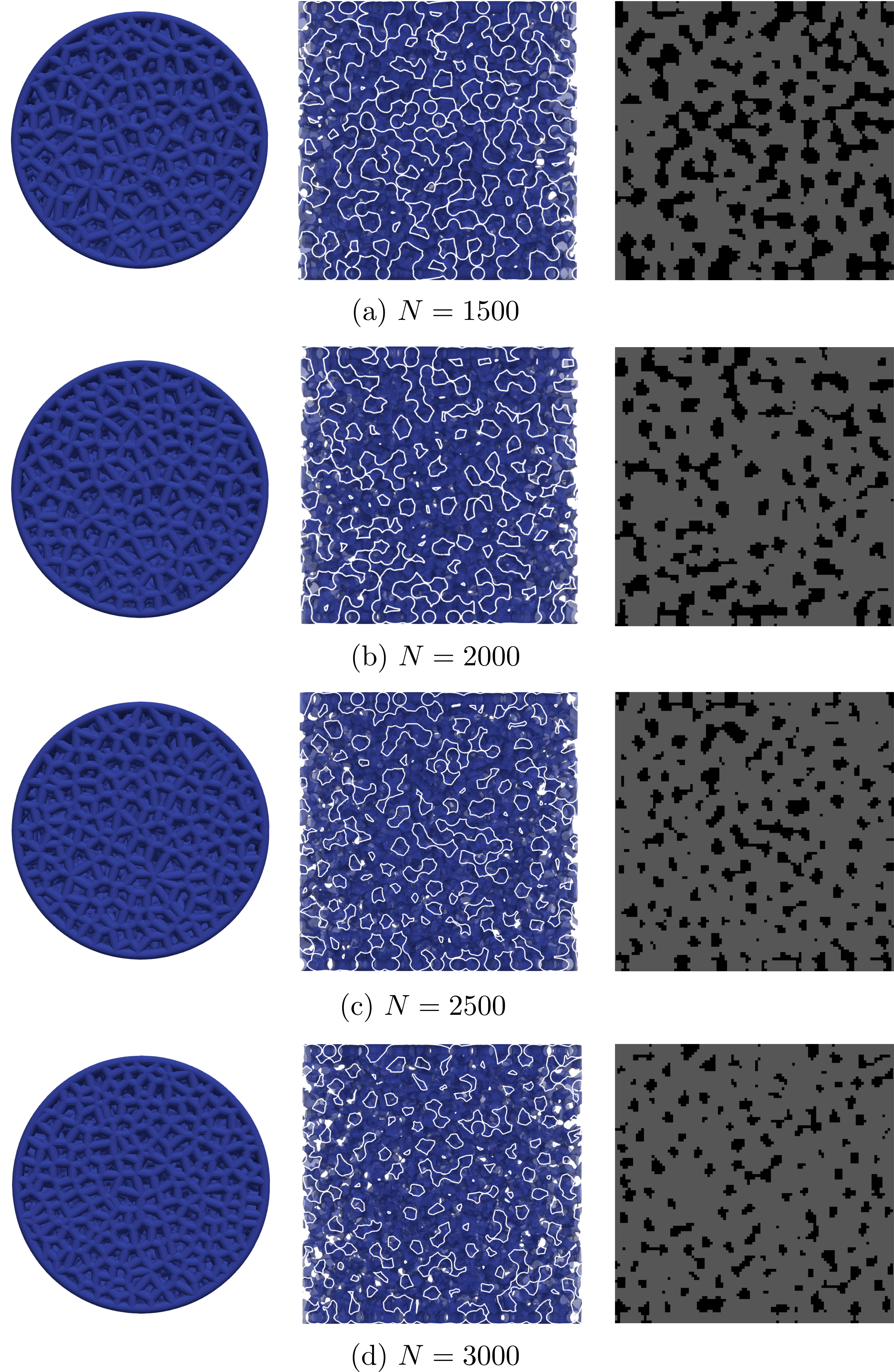}
    \caption{Resulting geometries for $N=1500,~2000,~2500,~3000$ seeds. First a top view of the inner porous network, in the middle, both, a 3D projection of a half view and central slice are provided to represent the porous network. And its corresponding voxelised representation is shown at the right-hand side. Only the confined porous network is presented.}
    \label{fig:3}
\end{figure}

The four geometries described in Section \ref{sec:geom} were generated. The volume of the domain envelope was 20,125.83 \si{\cubic\milli\meter}. This value was also used to estimate the porosity of the medium after the computation. The number of cylinders and spheres generated, the computation times, and porosity are provided in Table \ref{tab:1}. Figure \ref{fig: comp domain} shows an example of the entire STL geometry in its confined configuration used in simulations to ensure the flow direction in the chosen principal direction (co-linear to the principal axis of the cylinder), and Figure \ref{fig:3} shows the final network for each number of seeds. On the left-hand side, a clipped 3D-like image is presented within a 2D projection of the medial plane (in white) and its corresponding voxelised version for simulations at the right-hand side. The black voxels are void, whereas the grey voxels are the solid phase. 

\begin{table}[ht!]
\centering
\begin{tabular}{c|c|c|c|c}
 \begin{tabular}[c]{@{}c@{}}Seeds\\{[}~-~{]}\end{tabular} &
   \begin{tabular}[c]{@{}c@{}}Porosity \\ {[}~-~{]}\end{tabular} &
  \begin{tabular}[c]{@{}c@{}}Cylinders/spheres\\  {[}~-~{]}\end{tabular} &
  \begin{tabular}[c]{@{}c@{}}Time (CSG tree)\\  {[}~h.m.s~{]}\end{tabular} &
   \begin{tabular}[c]{@{}c@{}}Time (voxelising)\\  {[}~m.s~{]}\end{tabular}\\ \hline
1500  & 29.31\si{\percent}& 19714/656 & 08\si{\hour} 11\si{\minute}  &  09\si{\minute} 12\si{\second} \\
2000  &  22.40\si{\percent} & 26126/778 & 10\si{\hour} 37\si{\minute} &  13\si{\minute} 23\si{\second}\\
2500  & 17.43\si{\percent} & 32674/870 & 13\si{\hour} 05\si{\minute} &   16\si{\minute} 54\si{\second}\\
3000  & 10.13\si{\percent} & 37957/943 & 17\si{\hour} 46\si{\minute} &   22\si{\minute} 03\si{\second}\\
\end{tabular}
\caption{Geometrical and computational information of the generated synthetic samples. The voxelising time is provided for a required resolution of $N_x\times N_y \times N_z=80\times 80 \times 160$ voxels.}
\label{tab:1}
\end{table}

An increasing number of seeds resulted in an increasing number of cylinders/spheres. Therefore, more elements that interact all together were present in the CSG tree, leading to larger computation times. Depending on geometric choices and the number of seeds, the porosity of the scaffolds varied approximately from 10\% to 30 \%, as can also be observed in Figure \ref{fig:3}.

\newgeometry{top=2cm, bottom=2cm}
\restoregeometry

\subsection{Fluid/Structure interactions simulations}
In this section, only a single microstructure is used to present the simulation scenario, the one identified by the seeding number $N=3000$, since the main goal of this section is to serve as an application example for the methodology and no direct practical application is considered. The following is intended to present the general methodology without further analysis in the physical fields resulting from the simulation.

Here we consider the numerical implementation of the incompressible fluid flow simulations using previous section synthetic porous microstructures. The numerical implementation is achieved using the open-source C++ library \href{https://openfpm.mpi-cbg.de/}{OpenFPM}, which already integrates the DC-PSE framework.

Regarding the numerical setup for the simulations, first, since one of the goals is to estimate the macroscopic permeability of the medium according to \cite{whitaker1998method}, the non-slip condition on the fluid/structure of the interphase was defined, which represents the rigid behaviour of the solid phase. A test pressure of $9.93 $ Pa was defined as the reference pressure, and the controlled boundary condition for the parametric study was the inlet pressure, chosen so that the actual controlled parameter was the pressure difference between the inlet and outlet cross sections of the cylindrical geometry (see Fig. \ref{fig: comp domain}). The purpose of such a study is to observe the changes in the permeability of the sample for different Reynolds numbers as the velocity profile is affected by the imposed inlet pressure. The discrete interval of the imposed pressure difference for the parametric study was: 

\begin{equation*}
    \Delta p = \{ 5e-6, 1e-5, 5e-5, 1e-4, 5e-4, 1e-3, 5e-3, 1e-2, 5e-2, 1e-1, 5e-1, 1e0 \} [\text{Pa}]
\end{equation*}

\begin{figure}[!ht]
\centering
	\begin{subfigure}[t]{0.49\linewidth}
		\centering \includegraphics[scale=0.35]{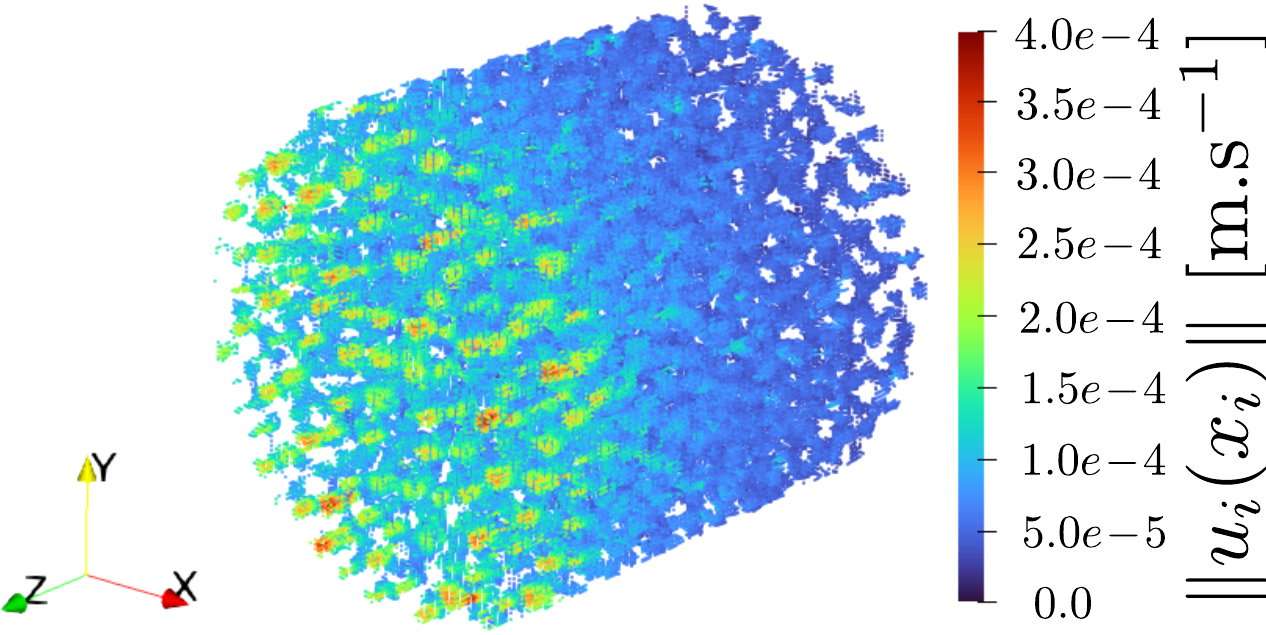}    
		\subcaption{{}}
		\label{fig: axial deformation}
	\end{subfigure} \hfill
	\begin{subfigure}[t]{0.49\linewidth}   
		\centering \includegraphics[scale=0.35]{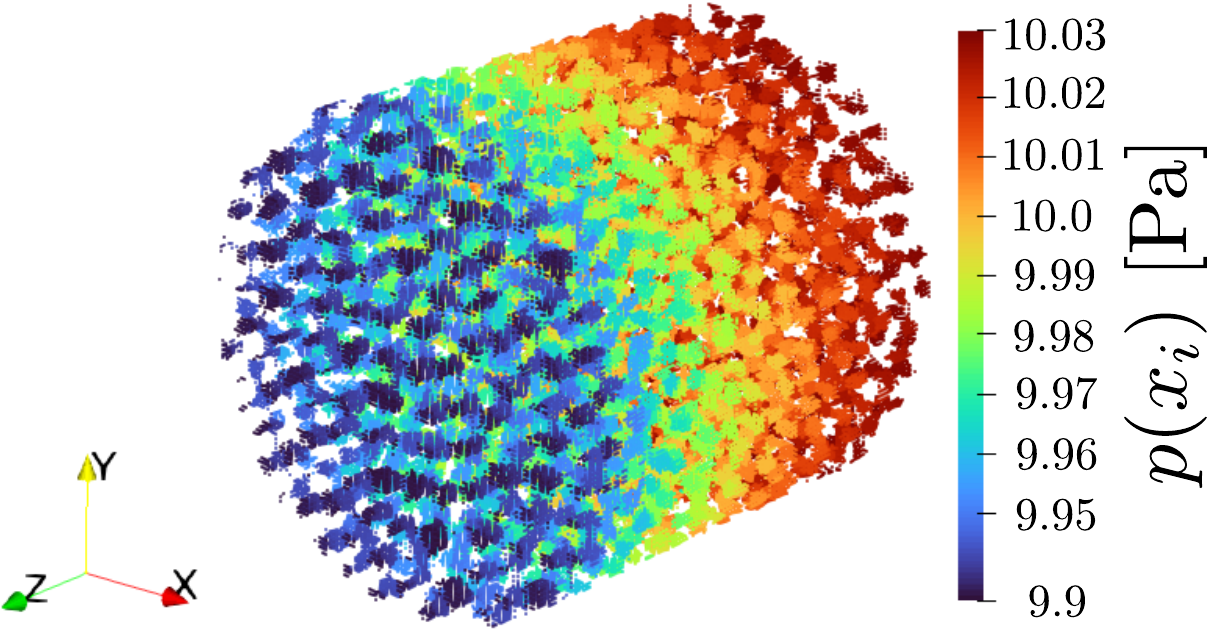}
		\subcaption{{}}
		\label{fig:tensile vs T}
	\end{subfigure}
	\caption{Example output fields R.O.I. (i.e. the porous network): (a) velocity norm distribution and (b) pressure distribution. For visualization only liquid subdomain is shown (i.e. $\chi(x_i) = 0$). }
	\label{fig: OPs fields}
\end{figure}

\begin{figure}[htp!]
\centering
        \includegraphics[width=\textwidth]{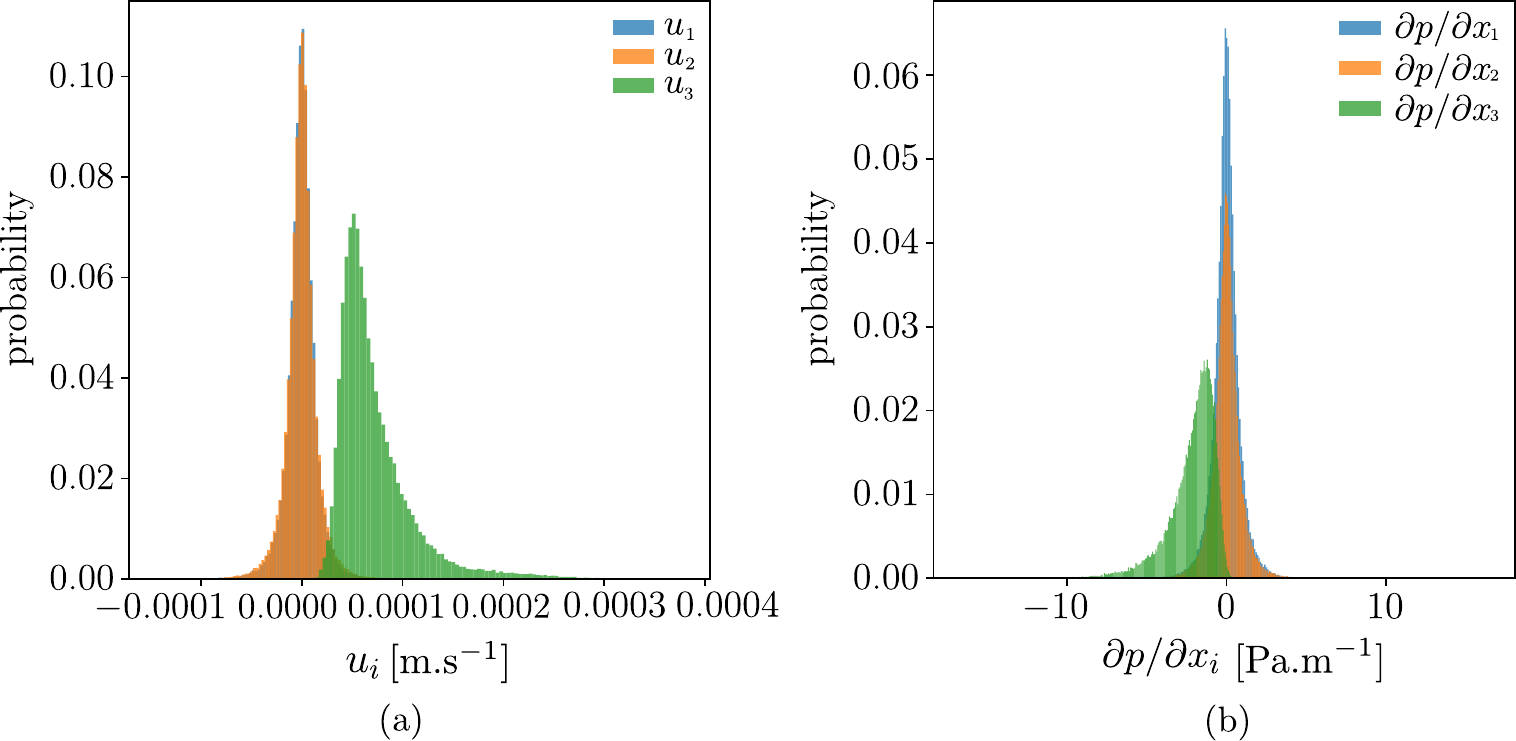}
        \caption{Example output fields R.O.I. (i.e. the porous network) Histograms : (a) velocity distribution and (b) pressure distribution.}
    \label{fig: OPs fields Histo}
\end{figure}

The output of the simulations are the velocity, $u_i(x_i)$, and pressure, $p(x_i)$ fields. This output data is the input that will be used to analyse the existence of a macroscopic permeability tensor ($\Tilde{\boldsymbol{K}}$), an example of such data for an input pressure $p = 10.03$ Pa (i.e. $\Delta p = 1e-2$ Pa) is presented in Fig. \ref{fig: OPs fields}, which only shows the physical fields for the fluid phase, and in the form of histograms in Fig. \ref{fig: OPs fields Histo}. The latter showing the principal direction of the fluid (i.e., $e_3$ or $z$ axis, histograms in green) where changes in pressure and velocity exhibit skewed distributions, while the rest of the components are represented by Gaussian-like distributions centred close to zero. In Fig. \ref{fig: OPs fields}, the norm of the velocity field show a non-linear evolution of the magnitude in the principal direction ($z$ axis) while the distributions in the other two principal directions appear more or less homogeneous for a given cross section. The pressure field, in contrast, exhibits a linear-like evolution from the higher input pressure to the lower outlet pressure.

\subsection{Permeability evaluation}

\subsubsection{Verification of theoretical hypotheses and criteria}

\begin{figure}[htp!]
\centering
        \includegraphics[width=\textwidth]{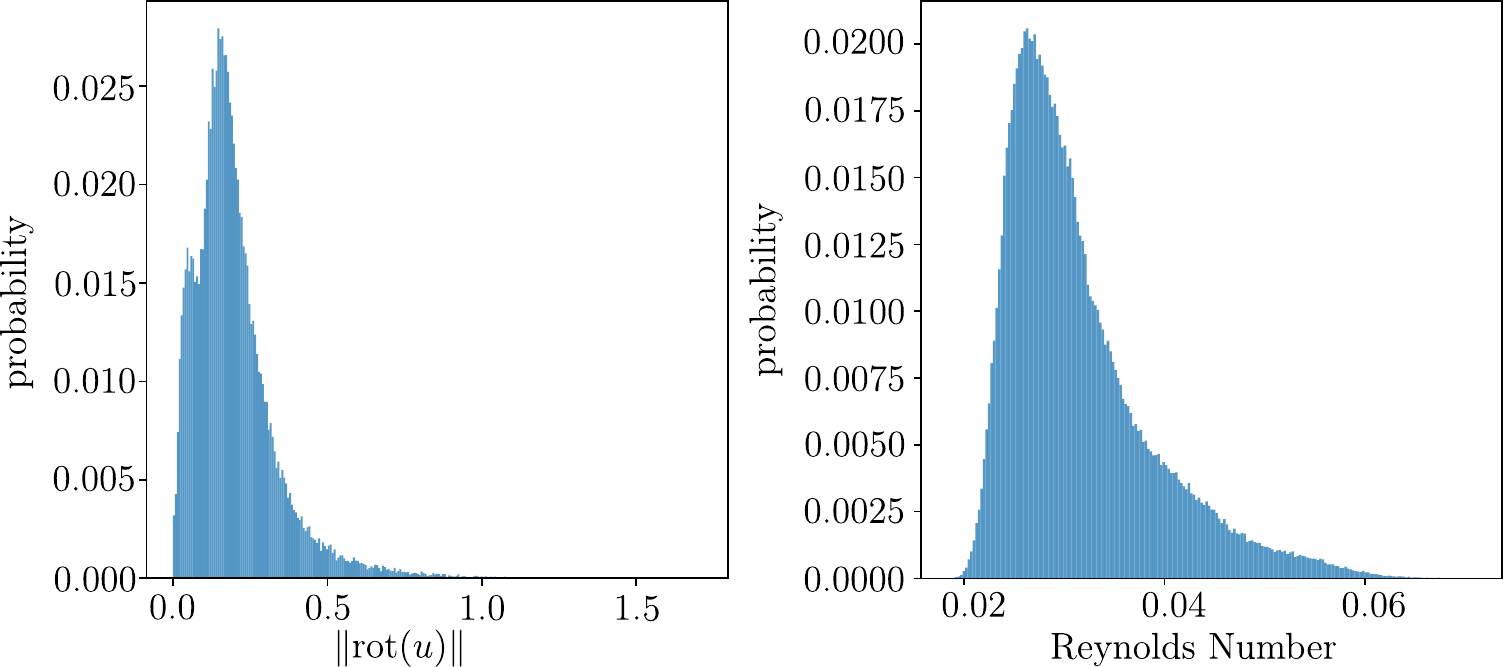}
        \caption{Distribution of creeping regime parameters for the example case, inlet pressure $p=10.003$ Pa. (a) The norm of the vorticity vector, and (b) The Reynolds number.}
	\label{fig: CreepingReg}
\end{figure}

\paragraph{Verification of creeping regime}
In the present work, the distribution of the norm of the vorticity vector in its original form is considered as a primary indicator of such regime and further verifications are given by the distribution of the main parameter of the creeping flow, the Reynolds number itself (see Fig. \ref{fig: CreepingReg}). This is because the methodology to compute the upscaling parameters is subject to arbitrary choices, including the spatial average and characteristic length within a representative elementary volume.

The figure above shows the evidence of the creeping regime for the reference example, the Reynolds number distribution follows the creeping regime condition $Re < 0.5$ \cite{Dybbs1984, Agnaou2017, Scandelli2022}. Both with the norm of the vorticity distribution ($\| \boldsymbol{w}(\boldsymbol{x}) \|$) suggest a linearity of the relationship between the macroscopic pressure gradient and the velocity, where low vorticity values can be used as arguments for low non-linear and localised effects of the viscous flow due to higher local velocity profiles.

\begin{figure}[ht!]
\centering
        \includegraphics[width=
        \textwidth]{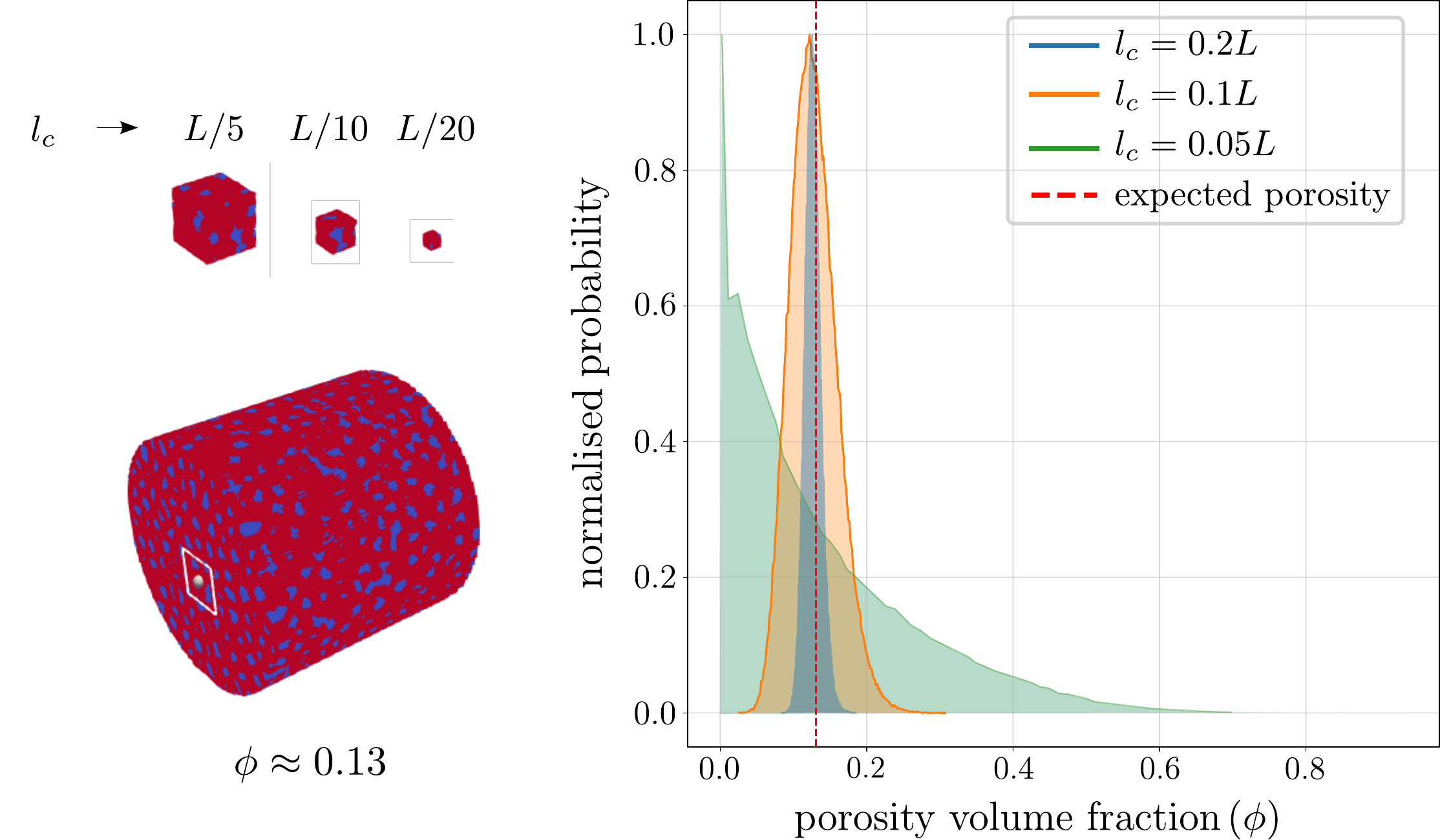}
        \caption{volume fraction distributions for sample characteristic lengths}
    \label{fig: REVs vfs}
\end{figure}

\paragraph{Identification of the characteristic length}
In Fig. \ref{fig: REVs vfs} one can observe the obtained volume fraction distributions for some examples of characteristic lengths, the magnitude of which is computed as a fraction of the total length of the macroscopic domain (i.e., $L$). As can be expected, as the characteristic length of the REV gets closer to the total length, the distribution starts to align with the expected volume fraction value (represented by the red dashed line). In order to maintain scale separation of the observed set of REVs, the maximum fraction of the total length is set to $l_c=0.2L$ with the mean volume fraction $\Bar{\phi} \simeq 0.1265$ (with $\Delta  \phi \simeq 0.004$ from the macroscopic volume fraction) and with the $70 \%$ of the data inside a symmetric extension of one standard deviation. 

\subsubsection{A One-dimension approximation for experimental comparisons}

\begin{figure}[ht!]
\centering
        \includegraphics[scale=0.5]{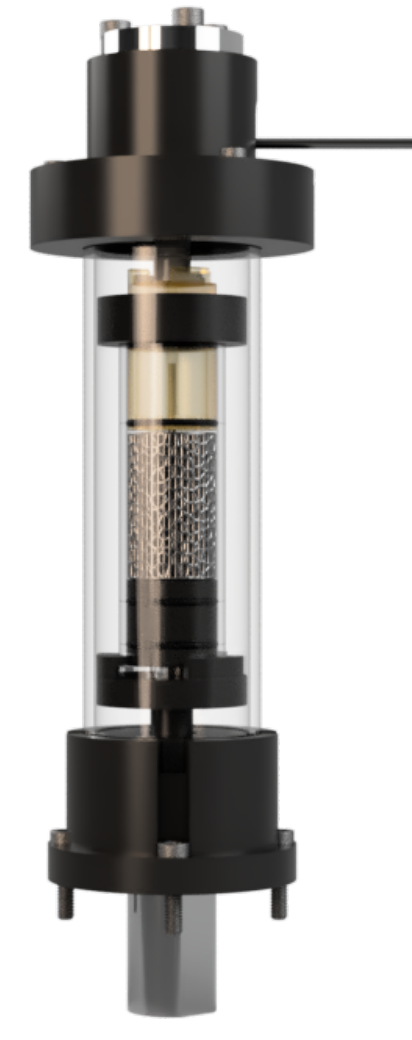}
        \caption{Example of an experimental setup compatible with an instrumented micro-CT scanner, for  permeability evaluation.}
    \label{fig: ExpSetup}
\end{figure}

Since this study
presents simulations to illustrate the capabilities of the proposed methodology, the three-dimensional permeability tensor is not explicitly addressed in the implementation, we propose here a one-dimensional approximation that serves as a link between simulations and actual experimental setups. As can be observed in Fig. \ref{fig: ExpSetup}, in a basic scenario the pressure gradient is only accessible for the direction which is collinear to the principal direction of the fluid flow (i.e. the $z$ direction from the conventions adopted previously); thus for a quasi-static established regime characterised by low Reynolds number, one is able to estimate the permeability of the porous medium by computing the pressure difference from input to output.

As mentioned above, the simulation scenario adopted in the present study is consistent with the criteria defined in \cite{Scandelli2022} for non-periodic microstructures; therefore, we can directly obtain a first approximation of the permeability in the $\boldsymbol{z}$ (or $\boldsymbol{e}_3$) direction (i.e. $\bar{\boldsymbol{K}}_{33}$) by assuming that the pressure field is constant at each infinitesimal cross section (i.e. $p(x_i) = p(x_3)$). 

\begin{equation}
    \bar{\boldsymbol{K}}_{33} \approx -\mu \phi \left(\frac{\Delta p}{\Delta z}\right)^{-1} \langle u_{3} \rangle,
\end{equation}

\begin{figure}[ht!]
\centering
        \includegraphics[width = \textwidth]{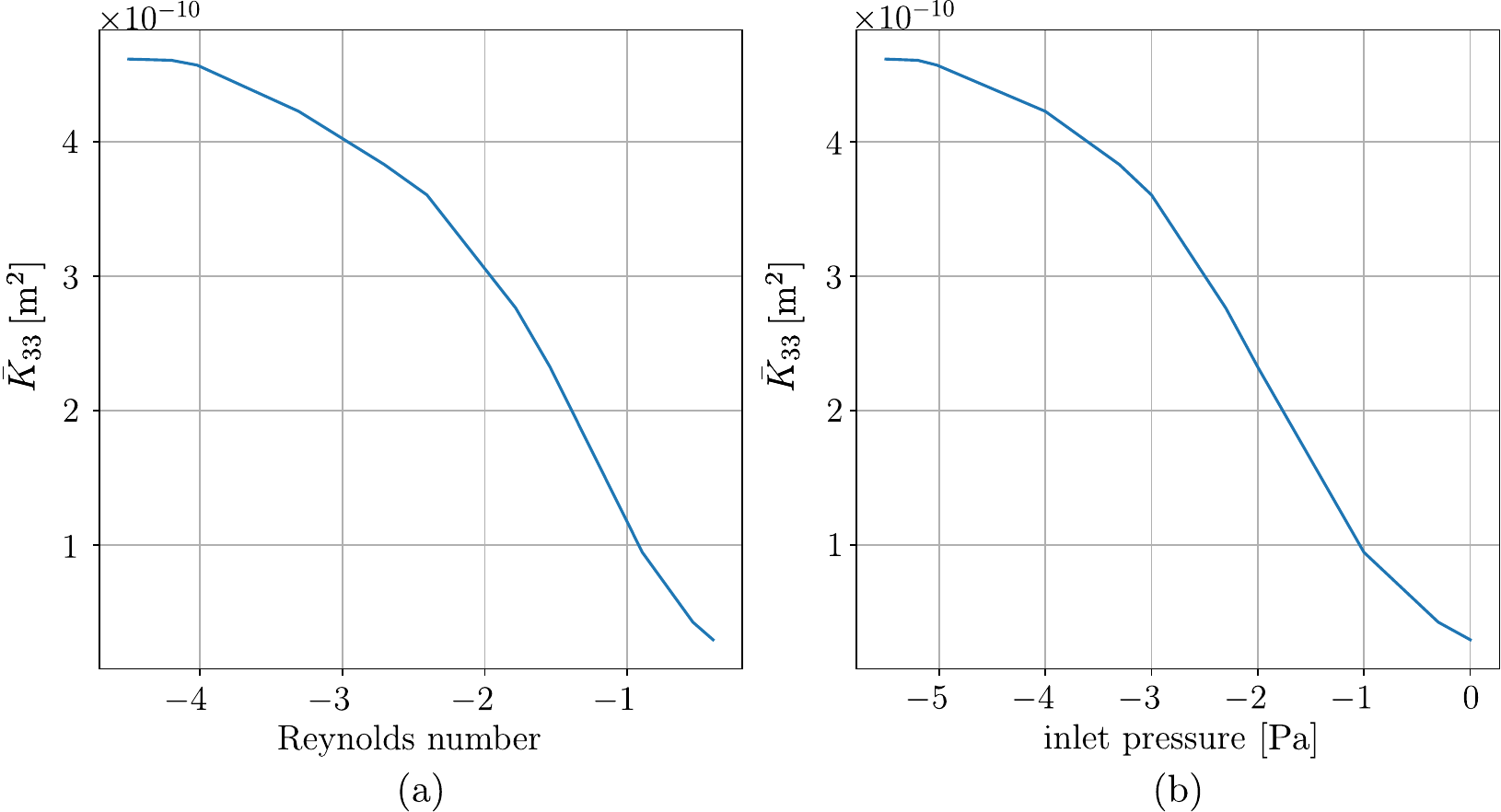}
        \caption{Evolution of the principal direction permeability $\bar{K}_{33}$, at the left side in function of the decimal logarithm of the Reynolds number and at the right-hand side, in function of the decimal logarithm of the imposed inlet pressure.}
    \label{fig: K33evol}
\end{figure}

Where $\Delta p$ represents the difference in pressure between the mean pressure of the inlet and outlet cross sections, $\Delta z$, the length of the porous medium under testing, and recalling from previous sections, $\langle u_3 \rangle$ the mean velocity of the fluid in the third direction (or the effective velocity in $z$ direction), $\mu$, the dynamic viscosity of the fluid, here the water (0.001 $\text{Pa} .\text{s}$), and $\phi$, the macroscopic volume fraction of porosity. The above equation is in fact equivalent to the earliest historical studies from \citet{darcy1856}. In Fig. \ref{fig: K33evol} we can observe the evolution of permeability as a function of the Reynolds number (indicator of the presence of the creeping regime) and inlet pressure (the controlled parameter of the subsequent simulations). In this figure, we can observe the transition from a linear-dominated regime to a more chaotic and turbulent one, as evidenced by the plateau-like behaviour of the left side (low Reynolds number zone), where the calculated parameter seems to be close to reaching a stable value. As the inlet pressure increases (as well as the Reynolds number) non-linear effects become more dominant and the Darcy-like assumption becomes invalid, which can be verified by the monotonic decrease on the right-hand side, suggesting that at higher pressure states the notion of permeability from a macroscopic Darcy law no longer holds. These observations are consistent with \cite{whitaker1998method, Scandelli2022}.

\section{Discussion}
\label{sec:discussion}

In this study, we have introduced a comprehensive methodology for generating synthetic porous media along with a direct computational pipeline that seamlessly integrates these structures into fluid-structure interaction simulations. This pipeline facilitates the transition from STL geometries to voxel-based domains and enables numerical simulations using the C++ library \href{https://openfpm.mpi-cbg.de/}{OpenFPM}. Through this framework, we have directly implemented the model described in Section \ref{sec: Math Modeling}. Furthermore, we have established a foundation for the analysis and estimation of the permeability of these synthetic samples, following the approaches of \citet{whitaker1998method, Whitaker1986}, and linking these estimates with the experimental evaluation. The ultimate goal is to provide parametric control over porous structures for real-world applications, such as prosthetics.

Using four different configurations of synthetic porous media (seed numbers: 1500, 2000, 2500, and 3000), the Constructive Solid Geometry (CSG) workflow generated structures with porosities ranging from approximately 10\% to 30\%. Higher seed numbers resulted in more intricate geometries, leading to increased computational times, with the 3000-seed case requiring nearly 18 hours for CSG processing.
Following voxelisation, numerical simulations using Entropically Damped Artificial Compressibility (EDAC) and discretisation-corrected particle strength exchange (DC-PSE) methods revealed that at low Reynolds numbers (Re < 0.5), the flow exhibited Darcy-like behaviour with negligible inertial effects. The estimated permeability plateaued under creeping flow conditions, demonstrating a linear relationship between the mean fluid velocity and the applied pressure gradient. However, as the inlet pressure increased, the non-linear effects became more pronounced, causing deviations from Darcy’s law. These findings highlight the robustness of the proposed pipeline in capturing different levels of porosity and the transition from creeping to more complex flow regimes within synthetic porous media.

Finally, we consider the full determination of the permeability tensor. This requires modifications to the generated microstructures to adjust the principal direction of fluid flow, along with refinements to the boundary condition to ensure consistency across successive simulations. Following the approaches outlined in \cite{Scandelli2022, pouya2002definition, Lang2014}, we propose to construct a linear system with a unique solution that allows the explicit formulation of the second-order effective permeability tensor for a given porous medium.

\clearpage

\section{Declaration of Competing Interest}
Authors have no conflicts of interest to report.

\textbf{Supplementary material:} The codes corresponding to the generation of the porous medium of this article are available at the following link: \url{https://github.com/Th0masLavigne/Porous_media_Generation_and_properties.git}.

\section{Acknowledgment}
 This research was funded in whole, or in part, by the Luxembourg National Research Fund (FNR),grant reference No. 17013182. For the purpose of open access, the author has applied a Creative Commons Attribution 4.0 International (CC BY 4.0) license to any Author Accepted Manuscript version arising from this submission.



\bibliographystyle{elsarticle-num-names} 
\bibliography{biblio,biblio_unlocked}

\begin{thebibliography}{46}
\expandafter\ifx\csname natexlab\endcsname\relax\def\natexlab#1{#1}\fi
\providecommand{\url}[1]{\texttt{#1}}
\providecommand{\href}[2]{#2}
\providecommand{\path}[1]{#1}
\providecommand{\DOIprefix}{doi:}
\providecommand{\ArXivprefix}{arXiv:}
\providecommand{\URLprefix}{URL: }
\providecommand{\Pubmedprefix}{pmid:}
\providecommand{\doi}[1]{\href{http://dx.doi.org/#1}{\path{#1}}}
\providecommand{\Pubmed}[1]{\href{pmid:#1}{\path{#1}}}
\providecommand{\bibinfo}[2]{#2}
\ifx\xfnm\relax \def\xfnm[#1]{\unskip,\space#1}\fi
\bibitem[{J.Gibson and M.~F(1999)}]{Gibson:1999}
\bibinfo{author}{L.~J.Gibson}, \bibinfo{author}{A.~M.~F},
\newblock \bibinfo{title}{Cellular {S}olids: {S}tructure and {P}roperties},
\newblock \bibinfo{journal}{Cambridge University Press}
  (\bibinfo{year}{1999}).
\bibitem[{Ashby et~al.(2000)Ashby, Evans, A., Gibson, Hutchinson, and
  and}]{Ashby:2000}
\bibinfo{author}{M.~Ashby}, \bibinfo{author}{A.~Evans},
  \bibinfo{author}{N.~A.}, \bibinfo{author}{L.~J. Gibson},
  \bibinfo{author}{J.~W. Hutchinson}, \bibinfo{author}{H.~J. G.~W. and},
\newblock \bibinfo{title}{Metal foams: A design guide,},
\newblock \bibinfo{journal}{Butterworth-Heinemann}  (\bibinfo{year}{2000}).
\bibitem[{Banhart(2001)}]{Banhart:2001}
\bibinfo{author}{J.~Banhart},
\newblock \bibinfo{title}{Manufacture, characterization and application of
  cellular metals and metal foams},
\newblock \bibinfo{journal}{Prog. Mater. Sci.} \bibinfo{volume}{46}
  (\bibinfo{year}{2001}) \bibinfo{pages}{559--632}.
\bibitem[{Boomsma and Poulikakos(2002)}]{Boomsma:2002}
\bibinfo{author}{K.~Boomsma}, \bibinfo{author}{D.~Poulikakos},
\newblock \bibinfo{title}{The effects of compression and pore size variations
  on the liquid flow characteristics of metal foams},
\newblock \bibinfo{journal}{Journal of Fluids Engineering}
  \bibinfo{volume}{124} (\bibinfo{year}{2002}) \bibinfo{pages}{263--272}.
\bibitem[{Lage et~al.(1996)Lage, Weinert, and C.Price}]{Lage:1996}
\bibinfo{author}{J.~L. Lage}, \bibinfo{author}{A.~Weinert},
  \bibinfo{author}{D.~C.Price},
\newblock \bibinfo{title}{Numerical study of a low permeability microporous
  heat sink for cooling phased-array radar systems},
\newblock \bibinfo{journal}{International Journal of Heat and Mass Transfer}
  \bibinfo{volume}{39} (\bibinfo{year}{1996}) \bibinfo{pages}{3633--3647}.
\bibitem[{Bhattacharya et~al.(2002)Bhattacharya, Calmidi, and
  Mahajan}]{Bhattacharya:2002}
\bibinfo{author}{A.~Bhattacharya}, \bibinfo{author}{V.~V. Calmidi},
  \bibinfo{author}{R.~L. Mahajan},
\newblock \bibinfo{title}{Thermophysical properties of high porosity metal
  foams},
\newblock \bibinfo{journal}{International Journal of Heat and Mass Transfer}
  \bibinfo{volume}{45} (\bibinfo{year}{2002}) \bibinfo{pages}{1017--1031}.
\bibitem[{Cowin(1985)}]{cowin1985}
\bibinfo{author}{S.~Cowin},
\newblock \bibinfo{title}{Mechanical properties of porous media},
\newblock \bibinfo{journal}{Journal of Biomechanics} \bibinfo{volume}{18}
  (\bibinfo{year}{1985}) \bibinfo{pages}{312--318}.
\bibitem[{Boutin and Auriault(1991)}]{boutin1991}
\bibinfo{author}{C.~Boutin}, \bibinfo{author}{J.~Auriault},
\newblock \bibinfo{title}{Dynamic behaviour of porous media saturated by a
  newtonian fluid: I. inertial effects},
\newblock \bibinfo{journal}{International Journal of Engineering Science}
  \bibinfo{volume}{29} (\bibinfo{year}{1991}) \bibinfo{pages}{81--93}.
\bibitem[{Biot(1956)}]{biot1956}
\bibinfo{author}{M.~Biot},
\newblock \bibinfo{title}{Theory of propagation of elastic waves in a
  fluid-saturated porous solid. i. low-frequency range},
\newblock \bibinfo{journal}{The Journal of the Acoustical Society of America}
  \bibinfo{volume}{28} (\bibinfo{year}{1956}) \bibinfo{pages}{168--178}.
\bibitem[{Bear(1972)}]{bear1972}
\bibinfo{author}{J.~Bear}, \bibinfo{title}{Dynamics of Fluids in Porous Media},
  \bibinfo{publisher}{American Elsevier}, \bibinfo{year}{1972}.
\bibitem[{Adler(1992)}]{adler1992}
\bibinfo{author}{P.~Adler}, \bibinfo{title}{Porous Media: Geometry and
  Transports}, \bibinfo{publisher}{Butterworth-Heinemann},
  \bibinfo{year}{1992}.
\bibitem[{Tien(1993)}]{Tien1993}
\bibinfo{author}{C.~Tien},
\newblock \bibinfo{title}{Principles of filtration},
\newblock \bibinfo{journal}{Chemical Engineering Progress} \bibinfo{volume}{89}
  (\bibinfo{year}{1993}) \bibinfo{pages}{49--58}.
\bibitem[{Bear(1972)}]{Bear:1972}
\bibinfo{author}{J.~Bear}, \bibinfo{title}{Dynamics of Fluids in Porous Media},
  \bibinfo{publisher}{American Elsevier Publishing Company},
  \bibinfo{year}{1972}.
\bibitem[{Allard and Atalla(2009)}]{Allard:2009}
\bibinfo{author}{J.~F. Allard}, \bibinfo{author}{N.~Atalla},
  \bibinfo{title}{Propagation of Sound in Porous Media: Modelling Sound
  Absorbing Materials}, \bibinfo{edition}{2} ed., \bibinfo{publisher}{John
  Wiley \& Sons}, \bibinfo{year}{2009}.
\bibitem[{Wu and Zhang(2016)}]{wu2016}
\bibinfo{author}{J.~Wu}, \bibinfo{author}{Y.~Zhang},
\newblock \bibinfo{title}{Hydrogels for 3d printing: Materials and
  applications},
\newblock \bibinfo{journal}{ACS Biomaterials Science \& Engineering}
  \bibinfo{volume}{3} (\bibinfo{year}{2016}) \bibinfo{pages}{1599--1611}.
\bibitem[{Martínez-Hernández et~al.(2007)Martínez-Hernández,
  Velasco-Santos, De-Icaza, and Castaño}]{martinez2007}
\bibinfo{author}{A.~Martínez-Hernández}, \bibinfo{author}{C.~Velasco-Santos},
  \bibinfo{author}{M.~De-Icaza}, \bibinfo{author}{V.~Castaño},
\newblock \bibinfo{title}{Microstructural characterisation of carbon nanotube
  composites},
\newblock \bibinfo{journal}{Composite Interfaces} \bibinfo{volume}{14}
  (\bibinfo{year}{2007}) \bibinfo{pages}{753--766}.
\bibitem[{Krishnan et~al.(2006)Krishnan, Murthy, and Garimella}]{Krishnan:2006}
\bibinfo{author}{S.~Krishnan}, \bibinfo{author}{J.~Y. Murthy},
  \bibinfo{author}{S.~V. Garimella},
\newblock \bibinfo{title}{Direct simulation of transport in open-cell metal
  foam},
\newblock \bibinfo{journal}{ASME J. Heat Transfer} \bibinfo{volume}{128}
  (\bibinfo{year}{2006}) \bibinfo{pages}{793--799}.
\bibitem[{du~Plessis et~al.(1994)du~Plessis, Montillet, Comiti, and
  Legrand}]{Plessis:1994}
\bibinfo{author}{J.~P. du~Plessis}, \bibinfo{author}{A.~Montillet},
  \bibinfo{author}{J.~Comiti}, \bibinfo{author}{J.~Legrand},
\newblock \bibinfo{title}{Pressure drop prediction for flow through high
  porosity metallic foams},
\newblock \bibinfo{journal}{Chem. Eng. Sci.} \bibinfo{volume}{49}
  (\bibinfo{year}{1994}) \bibinfo{pages}{3545--3553}.
\bibitem[{Boomsma and Poulikakos(2001)}]{Boomsma:2001}
\bibinfo{author}{K.~Boomsma}, \bibinfo{author}{D.~Poulikakos},
\newblock \bibinfo{title}{On the effective thermal conductivity of a
  three-dimensionally structured fluid-saturated metal foam},
\newblock \bibinfo{journal}{Int. J. Heat Mass Transfer} \bibinfo{volume}{44}
  (\bibinfo{year}{2001}) \bibinfo{pages}{827--836}.
\bibitem[{Bai et~al.(2012)Bai, Sun, Lin, Kennedy, and Williams}]{Bai:2011}
\bibinfo{author}{Y.~Bai}, \bibinfo{author}{D.~Sun}, \bibinfo{author}{J.~Lin},
  \bibinfo{author}{D.~Kennedy}, \bibinfo{author}{F.~Williams},
\newblock \bibinfo{title}{Numerical aerodynamic simulations of a {NACA} airfoil
  using {CFD} with block-iterative coupling and turbulence modelling},
\newblock \bibinfo{journal}{International Journal of Computational Fluid
  Dynamics} \bibinfo{volume}{26} (\bibinfo{year}{2012})
  \bibinfo{pages}{119--132}.
\bibitem[{Hil(1963)}]{Hill:1963}
\bibinfo{author}{R.~Hil},
\newblock \bibinfo{title}{Elastic properties of reinforced solids: some
  theoretical principles},
\newblock \bibinfo{journal}{Journal of the Mechanics and Physics of Solids}
  \bibinfo{volume}{11} (\bibinfo{year}{1963}) \bibinfo{pages}{357--372}.
\bibitem[{Aarnes and Efendiev(2008)}]{Aarnes:2008}
\bibinfo{author}{J.~E. Aarnes}, \bibinfo{author}{Y.~Efendiev},
\newblock \bibinfo{title}{Mixed multiscale finite element methods for
  stochastic {P}orous media flows},
\newblock \bibinfo{journal}{SIAM Journal on Scientific Computing}
  \bibinfo{volume}{30} (\bibinfo{year}{2008}) \bibinfo{pages}{2319--2339}.
\bibitem[{Ganapathysubramanian and Zabaras(2009)}]{Ganapathysubramanian:2009}
\bibinfo{author}{B.~Ganapathysubramanian}, \bibinfo{author}{N.~Zabaras},
\newblock \bibinfo{title}{A stochastic multiscale framework for modeling flow
  through random heterogeneous porous media},
\newblock \bibinfo{journal}{Journal of Computational Physics}
  \bibinfo{volume}{228} (\bibinfo{year}{2009}) \bibinfo{pages}{591--618}.
\bibitem[{Biswal et~al.(2011)Biswal, Oren, Held, Bakke, and
  Hilfer}]{Biswal:2011}
\bibinfo{author}{B.~Biswal}, \bibinfo{author}{P.-E. Oren},
  \bibinfo{author}{R.~J. Held}, \bibinfo{author}{S.~Bakke},
  \bibinfo{author}{R.~Hilfer},
\newblock \bibinfo{title}{Modeling of multiscale porous media},
\newblock \bibinfo{journal}{Image Analysis \& Stereology} \bibinfo{volume}{28}
  (\bibinfo{year}{2011}).
\bibitem[{Vazic et~al.(2022)Vazic, Abali, Yang et~al.}]{Vasic:2022}
\bibinfo{author}{B.~Vazic}, \bibinfo{author}{B.~Abali},
  \bibinfo{author}{H.~Yang}, et~al.,
\newblock \bibinfo{title}{Mechanical analysis of heterogeneous materials with
  higher-order parameters},
\newblock \bibinfo{journal}{Engineering with Computers} \bibinfo{volume}{38}
  (\bibinfo{year}{2022}) \bibinfo{pages}{5051--5067}.
  \DOIprefix\doi{10.1007/s00366-021-01555-9}.
\bibitem[{Dostert et~al.(2008)Dostert, Efendiev, and Hou}]{Dostert:2008}
\bibinfo{author}{P.~Dostert}, \bibinfo{author}{Y.~Efendiev},
  \bibinfo{author}{T.~Hou},
\newblock \bibinfo{title}{Multiscale finite element methods for stochastic
  porous media flow equations and application to uncertainty quantification},
\newblock \bibinfo{journal}{Computer Methods In Applied Mechanics and
  Engineering} \bibinfo{volume}{197} (\bibinfo{year}{2008})
  \bibinfo{pages}{3445--3455}.
\bibitem[{Whitaker(1998)}]{whitaker1998method}
\bibinfo{author}{S.~Whitaker}, \bibinfo{title}{The method of volume averaging},
  volume~\bibinfo{volume}{13}, \bibinfo{publisher}{Springer Science \& Business
  Media}, \bibinfo{year}{1998}.
\bibitem[{Scandelli et~al.(2022)Scandelli, Ahmadi-Senichault, Levet, and
  Lachaud}]{Scandelli2022}
\bibinfo{author}{H.~Scandelli}, \bibinfo{author}{A.~Ahmadi-Senichault},
  \bibinfo{author}{C.~Levet}, \bibinfo{author}{J.~Lachaud},
\newblock \bibinfo{title}{Computation of the permeability tensor of
  non-periodic anisotropic porous media from 3d images},
\newblock \bibinfo{journal}{Transport in Porous Media} \bibinfo{volume}{142}
  (\bibinfo{year}{2022}) \bibinfo{pages}{669--697}. \URLprefix
  \url{https://link.springer.com/article/10.1007/s11242-022-01766-8}.
  \DOIprefix\doi{10.1007/S11242-022-01766-8/TABLES/6}.
\bibitem[{Quey et~al.(2011)Quey, Dawson, and Barbe}]{Quey2011}
\bibinfo{author}{R.~Quey}, \bibinfo{author}{P.~Dawson},
  \bibinfo{author}{F.~Barbe},
\newblock \bibinfo{title}{Large-scale 3d random polycrystals for the finite
  element method: Generation, meshing and remeshing},
\newblock \bibinfo{journal}{Computer Methods in Applied Mechanics and
  Engineering} \bibinfo{volume}{200} (\bibinfo{year}{2011})
  \bibinfo{pages}{1729--1745}. \URLprefix
  \url{https://doi.org/10.1016/j.cma.2011.01.002}.
  \DOIprefix\doi{10.1016/j.cma.2011.01.002}.
\bibitem[{Zhou(2019)}]{zhou2019pymesh}
\bibinfo{author}{Q.~Zhou},
\newblock \bibinfo{title}{Pymesh—geometry processing library for python},
\newblock \bibinfo{journal}{Software available for download at https://github.
  com/PyMesh/PyMesh}  (\bibinfo{year}{2019}).
\bibitem[{Pedrekoff(2015)}]{Pederkoff2015}
\bibinfo{author}{C.~Pedrekoff}, \bibinfo{title}{stl-to-voxel python library},
  \bibinfo{howpublished}{\url{https://github.com/cpederkoff/stl-to-voxel/blob/master/LICENCE.md}},
  \bibinfo{year}{2015}.
\bibitem[{Singh et~al.(2023)Singh, Sbalzarini, and Obeidat}]{Singh:2023}
\bibinfo{author}{A.~Singh}, \bibinfo{author}{I.~F. Sbalzarini},
  \bibinfo{author}{A.~Obeidat},
\newblock \bibinfo{title}{Entropically {D}amped {A}rtificial {C}ompressibility
  for the {D}iscretization {C}orrected {P}article {S}trength {E}xchange
  {M}ethod in {I}ncompressible {F}luid {M}echanics},
\newblock \bibinfo{journal}{Computers \& Fluids} \bibinfo{volume}{267}
  (\bibinfo{year}{2023}) \bibinfo{pages}{106074}.
  \DOIprefix\doi{https://doi.org/10.1016/j.compfluid.2023.106074}.
\bibitem[{Clausen(2013)}]{Clausen:2013}
\bibinfo{author}{J.~R. Clausen},
\newblock \bibinfo{title}{Entropically damped form of artificial
  compressibility for explicit simulation of incompressible flow},
\newblock \bibinfo{journal}{Phys.\ Rev.} \bibinfo{volume}{87}
  (\bibinfo{year}{2013}) \bibinfo{pages}{013309}.
\bibitem[{Obeidat and Bordas(2019)}]{Obeidat:2019}
\bibinfo{author}{A.~Obeidat}, \bibinfo{author}{S.~P.~A. Bordas},
\newblock \bibinfo{title}{An implicit boundary approach for viscous
  compressible high {R}eynolds flows using a hybrid remeshed particle
  hydrodynamics method},
\newblock \bibinfo{journal}{Journal of Computational Physics}
  \bibinfo{volume}{391} (\bibinfo{year}{2019}) \bibinfo{pages}{347--364}.
\bibitem[{Obeidat et~al.(2020)Obeidat, Andreas, Bordas, and
  Ziliana}]{Obeidat:2020}
\bibinfo{author}{A.~Obeidat}, \bibinfo{author}{T.~Andreas},
  \bibinfo{author}{S.~P.~A. Bordas}, \bibinfo{author}{A.~Ziliana},
\newblock \bibinfo{title}{Discrete filters for viscous compressible high
  reynolds flows in industrial complex geometry using a hybrid remeshed
  particle hydrodynamics method}  (\bibinfo{year}{2020}).
\bibitem[{Schrader et~al.(2010)Schrader, Reboux, and
  Sbalzarini}]{Schrader:2010}
\bibinfo{author}{B.~Schrader}, \bibinfo{author}{S.~Reboux},
  \bibinfo{author}{I.~F. Sbalzarini},
\newblock \bibinfo{title}{Discretization correction of general integral {PSE}
  operators for particle methods},
\newblock \bibinfo{journal}{Journal of Computational Physics}
  \bibinfo{volume}{229} (\bibinfo{year}{2010}) \bibinfo{pages}{4159--4182}.
\bibitem[{Degond and Mas-Gallic(1989)}]{Degond:1989a}
\bibinfo{author}{P.~Degond}, \bibinfo{author}{S.~Mas-Gallic},
\newblock \bibinfo{title}{The weighted particle method for convection-diffusion
  equations. part 1: The case of an isotropic viscosity},
\newblock \bibinfo{journal}{Mathematics of Computation} \bibinfo{volume}{53}
  (\bibinfo{year}{1989}) \bibinfo{pages}{485--507}. \bibinfo{note}{Oct}.
\bibitem[{Eldredge et~al.(2002)Eldredge, Leonard, and Colonius}]{Eldredge:2002}
\bibinfo{author}{J.~D. Eldredge}, \bibinfo{author}{A.~Leonard},
  \bibinfo{author}{T.~Colonius},
\newblock \bibinfo{title}{A general determistic treatment of derivatives in
  particle methods},
\newblock \bibinfo{journal}{Journal of Computational Physics}
  \bibinfo{volume}{180} (\bibinfo{year}{2002}) \bibinfo{pages}{686--709}.
\bibitem[{Howes and Whitaker(1985)}]{Howes1985}
\bibinfo{author}{F.~A. Howes}, \bibinfo{author}{S.~Whitaker},
\newblock \bibinfo{title}{The spatial averaging theorem revisited},
\newblock \bibinfo{journal}{Chemical Engineering Science} \bibinfo{volume}{40}
  (\bibinfo{year}{1985}) \bibinfo{pages}{1387--1392}.
  \DOIprefix\doi{10.1016/0009-2509(85)80078-6}.
\bibitem[{Whitaker(1986)}]{Whitaker1986}
\bibinfo{author}{S.~Whitaker},
\newblock \bibinfo{title}{Flow in porous media i: A theoretical derivation of
  darcy's law},
\newblock \bibinfo{journal}{Transport in Porous Media} \bibinfo{volume}{1}
  (\bibinfo{year}{1986}) \bibinfo{pages}{3--25}. \URLprefix
  \url{https://link.springer.com/article/10.1007/BF01036523}.
  \DOIprefix\doi{10.1007/BF01036523/METRICS}.
\bibitem[{Pouya and Courtois(2002{\natexlab{a}})}]{pouya2002definition}
\bibinfo{author}{A.~Pouya}, \bibinfo{author}{A.~Courtois},
\newblock \bibinfo{title}{Definition of the permeability of fractured rock
  masses by homogenisation methods.},
\newblock \bibinfo{journal}{Comptes Rendus Geoscience} \bibinfo{volume}{334}
  (\bibinfo{year}{2002}{\natexlab{a}}) \bibinfo{pages}{975--979}.
\bibitem[{Pouya and Courtois(2002{\natexlab{b}})}]{POUYA2002975}
\bibinfo{author}{A.~Pouya}, \bibinfo{author}{A.~Courtois},
\newblock \bibinfo{title}{Définition de la perméabilité équivalente des
  massifs fracturés par des méthodes d'homogénéisation},
\newblock \bibinfo{journal}{Comptes Rendus Geoscience} \bibinfo{volume}{334}
  (\bibinfo{year}{2002}{\natexlab{b}}) \bibinfo{pages}{975--979}. \URLprefix
  \url{https://www.sciencedirect.com/science/article/pii/S1631071302018394}.
  \DOIprefix\doi{https://doi.org/10.1016/S1631-0713(02)01839-4}.
\bibitem[{Lang et~al.(2014)Lang, Paluszny, and Zimmerman}]{Lang2014}
\bibinfo{author}{P.~S. Lang}, \bibinfo{author}{A.~Paluszny},
  \bibinfo{author}{R.~W. Zimmerman},
\newblock \bibinfo{title}{Permeability tensor of three-dimensional fractured
  porous rock and a comparison to trace map predictions},
\newblock \bibinfo{journal}{Journal of Geophysical Research: Solid Earth}
  \bibinfo{volume}{119} (\bibinfo{year}{2014}) \bibinfo{pages}{6288--6307}.
  \URLprefix
  \url{https://agupubs.onlinelibrary.wiley.com/doi/abs/10.1002/2014JB011027}.
  \DOIprefix\doi{https://doi.org/10.1002/2014JB011027}.
  \href{http://arxiv.org/abs/https://agupubs.onlinelibrary.wiley.com/doi/pdf/10.1002/2014JB011027}{{\tt
  arXiv:https://agupubs.onlinelibrary.wiley.com/doi/pdf/10.1002/2014JB011027}}.
\bibitem[{Dybbs and Edwards(1984)}]{Dybbs1984}
\bibinfo{author}{A.~Dybbs}, \bibinfo{author}{R.~V. Edwards},
\newblock \bibinfo{title}{A new look at porous media fluid mechanics — darcy
  to turbulent},
\newblock \bibinfo{journal}{NATO ASI Series, Series E: Applied Sciences}
  (\bibinfo{year}{1984}) \bibinfo{pages}{199--256}. \URLprefix
  \url{https://link.springer.com/chapter/10.1007/978-94-009-6175-3_4}.
  \DOIprefix\doi{10.1007/978-94-009-6175-3_4}.
\bibitem[{Agnaou et~al.(2017)Agnaou, Lasseux, and Ahmadi}]{Agnaou2017}
\bibinfo{author}{M.~Agnaou}, \bibinfo{author}{D.~Lasseux},
  \bibinfo{author}{A.~Ahmadi},
\newblock \bibinfo{title}{Origin of the inertial deviation from darcy's law: An
  investigation from a microscopic flow analysis on two-dimensional model
  structures},
\newblock \bibinfo{journal}{Physical Review E} \bibinfo{volume}{96}
  (\bibinfo{year}{2017}) \bibinfo{pages}{43105}. \URLprefix
  \url{https://hal.science/hal-03140925v1}.
  \DOIprefix\doi{10.1103/PhysRevE.96.043105ï}.
\bibitem[{Darcy(1856)}]{darcy1856}
\bibinfo{author}{H.~Darcy}, \bibinfo{title}{Les fontaines publiques de Dijon},
  \bibinfo{year}{1856}. \URLprefix
  \url{https://books.google.lu/books?hl=en&lr=&id=yXKx1zPVQMUC&oi=fnd&pg=PA4&dq=+Les+fontaines+publiques+de+la+ville+de+Dijon&ots=UdSn6lJdda&sig=8eDS1cds36LYizj-RMhXMpVu1g8&redir_esc=y#v=onepage&q=Les%20fontaines%20publiques%20de%20la%20ville%20de%20Dijon&f=false}.

\end{thebibliography}

\end{document}